# A computer-based recursion algorithm for automatic charge of power device of electric vehicles carrying electromagnet

Daniel Liu, Lvette Lopez*

Department of Electrical and Electronic Engineering, Hong Kong University of Science and Technology, Clear Water Bay, Kowloon, Hong Kong
*Corresponding author: lopez1@ust.hk

*Abstract*—This paper proposes a computer-based recursion algorithm for automatic charge of power device of electric vehicles carrying electromagnet. The charging system includes charging cable with one end connecting gang socket, electromagnetic gear driving the connecting socket and a charging pile breaking or closing, and detecting part for detecting electric vehicle static call or start state. The gang socket mentioned above is linked to electromagnetic gear, and the detecting part is connected with charging management system containing the intelligent charging power module which controls the electromagnetic drive action to close socket with a charging pile at static state and to break at start state. Our work holds an electric automobile with convenience, safety low maintenance cost.

## I. INTRODUCTION

Electric vehicle plays an increasingly important role in traffic and travel in recent years [1-3]. Large scale of lease business and project bid for charging station construction have been started one after another in metropolis like Shanghai, Beijing. Furthermore, self-service charging pile for electric vehicle users has been in batch construction near residential areas, parking and commercial district in major and medium city [4-9]. The above trend indicates that the development and popularization of electric vehicles in the nationwide have been an urgent building task in the current traffic industry development, which is also rather important for reducing dependence on foreign oil, protecting of national energy security, and realizing the sustainable development of economy and society. Up to now, state operation management system for electric vehicle is still battery pit change mode [10]. However, with the growing number of electric vehicle users, especially under the situation in which users aren't comfortably accessible to the power plant, the problem that electric vehicle with low battery can't get effective charging is more significant. Therefore, to solve above problem, state is starting to build a certain size of charging pile system in public parks, residential areas etc. As these pile systems are mostly of the self-help charging payment management mode, the resistance to wear of human-machine operation interface and a charging device are crucial for the system safety [11]. As a key factor in the charging process loss, users often forget to pull the charging cable out of the socket after charging and starting car. This mistake can directly cause physical damage on socket or cable plug, and then increase the self-help charging pile repair and operation cost. A number of electric vehicle manufacturers provide technique improvement to solve the problem. The company of Hall (UK) installs pressure sensitive resistance sheet at one end of the cable socket [12]. When users start car without pulling out the charging cable, the tensile force of vehicle movement will lead to the resistance value increasing rapidly in a short time, accordingly causing the charging current below the system default charging current threshold and finishing the charging process by automatic recognition of charging system [13-18]. This proposal can effectively ensure the safety of charging cable and socket. FAW-Volkswagen Automobile Co. Ltd places an alarm chip in the charging cable socket, which can generate alarm and remind users when cable is under tensile stress as user's misoperation.

Therefore, in order to effectively guarantee the safety of electric vehicle charging and simultaneously reduce the loss and avoid failure of charging pile caused by misoperation during charging or power off process, it is very necessary to build a perfect, safety, convenient charging pile interface protection device which can dramatically reduce the failure probability and cost of the maintenance and repair. Generally speaking, the main contents in this paper relates to the technology of electric vehicle charging, particularly to an electric automobile carrying automatic charging power device.

## II. THE TECHNICAL PROPOSAL TO ELECTRIC CHARGING POWER DEVICE

To overcome the deficiency of operation trouble, lower safety and high maintenance cost of electric vehicle charging interface, our paper intents to provide an electric vehicle carrying electromagnet automatic charging power device with convenience, safety, low maintenance cost.

For this purpose, we specifically offers the following technical proposal: firstly, it is assumed that an electric automobile carrying electromagnet automatic charging power device, including charging cable with one end connecting gang socket. The electric vehicle also includes an electromagnetic gear driving the connecting socket and a charging pile breaking or closing, and a detecting part for electric vehicle static call or start state. The gang socket said above is coupled with electromagnetic gear, and the detecting part is connected with charging management system containing the intelligent charging power module which controls the electromagnetic drive action to close socket with a charging pile at static state and to break at start state.

Furthermore, there is electromagnet embedded in the charging socket of charging pile, and metal thin layer adsorbing electromagnet covered on the gang socket. Secondly, the charging cable is covered with rubber insulation layer, and the electric contact of cable and socket are fixed at the edge of the rubber insulation layer respectively; the charging socket on charging pile is covered with rubber insulation layer, electric contact in that is connected with the two electric contact in the charging socket by the electric pump respectively. The novelty of our technical proposal is: a set of detection module, which can discern the car state and send charging or shutting command through charging management system; a connecting cable to realize charging function, one side of which is the gang socket connecting with charging device; electrical contact in cable and socket composed of narrow metal sheet, whose interval is adjustable. To solve this problem we will use Li et al. charging method initially introduced in their patent [19]. Their method indicated that when electric car is in static state, the detection module discerns the vehicle state, stores it for a stop command and sends it to the charging management system [19]. Based on Li's method, we will further design our system by using electromagnetic conversion device. Under the magnetic force, adsorption interaction will occur between cable and socket contact metal sheet, causing the socket and cable port on-state, and as a result, the vehicle is charged successfully.

On the other hand, when the electric car from static enters into the start state, the detection module discerns the vehicle state, stores it for a startup command and sends it to the charging management system [19]. Also by using such mechanism, our system adjusts the pole distribution of electric contact metal sheet by electromagnetic conversion device, causing anisotropic repulsion of cable with socket and at last loosing the connection between the two [20]. In this process, cable socket will rotate around its axis counterclockwise to the repulsive force parallel position by magnetic force; as a result, the cable is departed from the charging socket.

### III. APPROACH FOR THE DESIGN OF THE ELECTROMAGNET CHARGING DEVICE

In this section, we present the detailed description for the embodiment of the design of the charging device with drawings. Referring to figure 1-4, we provides an electric automobile carrying electromagnet automatic charging power device, including charging cable with one end connecting gang socket. The electric vehicle also includes an electromagnetic gear driving the connecting socket and a charging pile breaking or closing, and a detecting part for electric vehicle static call or start state. According to Li's description in their patent [19], we further develop their method by using the gang socket which coupled with electromagnetic gear, and the detecting part is connected with charging management system containing the intelligent charging power module which controls the electromagnetic drive action to close socket with a charging pile at static state and to break at start state.

As shown in Fig.1, the electric vehicle overlooking section structure (100) and the charging system contains battery (110) providing energy for electric driving device (115), electromagnetic gear (120). Battery can be charged through charging management system (125), which can obtain external energy in a variety of ways. The mode of (125) depends on the user's requirement. In the whole structure, battery (110) is connected with charging cable through socket connection device. (130) is the charging socket in the front of vehicle, the sockets at driver seat side are (131), (132), (133), and sockets at back seat side are (134), (135), (136). Vehicle can be connected with charging pile through sockets at different locations, and the concrete connecting mode can be selected according to the requirements of users. From above, we consider that the electric vehicle in our design procedure is the one with charging socket of multi port. Charging cable (140) can receive and distinguish from DC or AC electric signal.

Besides, the location selection of the charging cable port and the vehicle start running direction are one-one correspondence. For example, when the charging cable joins in socket 130, and users trigger the car forgetting to pull the charging cable out, 140 will be natural loss from socket 130. In reverse, when the cable is inserted into socket 134, the reverse running of vehicle will cause physical damage on charging cable or socket. All the access mode of charging cable can be selected flexibly according to user's demand. In practice, one or more charging socket can be used for preventing damage of charging socket or cable.

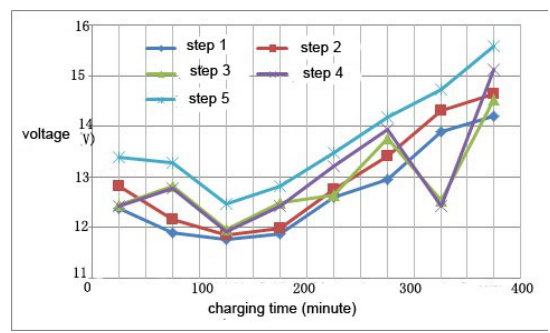

Fig. 1 The battery terminal voltage varying with time when the external current is input into the battery series in the electric vehicle.

Based on above implementation method, car charging management device contains a kind of electric socket 205 using for cable connection. Fig.2 shows the schematic diagram of electric socket with dentate joint, which includes a kind of electromagnetic induction device. The electromagnetic induction device can be joined to the automatic controller 150. In such aspect of work, we also develop he main function of the electromagnetic induction device which was shown by Li et al. in their latest patent [29]. We do this task because we intend to identify the connect state of charging cable. When vehicle starts, the induction device will real-time monitor the charging state and send visual instruction information to user interface with the sentence "The car enters into the non-starting state, and charging cable is disconnected". By sending this instruction, part damage caused by the neglect of pulling the charging cable out while starting to move the car can be avoided. In the present embodiment, induction device 210 real-time monitors all the sockets at different location in the charging system. On the basis of the above implementation method, charging socket is with electromagnet (220). This kind of electromagnet can magnetically absorb the electrical contact metal sheet in socket. Fig.2 reveals the rectangular electromagnet in socket.

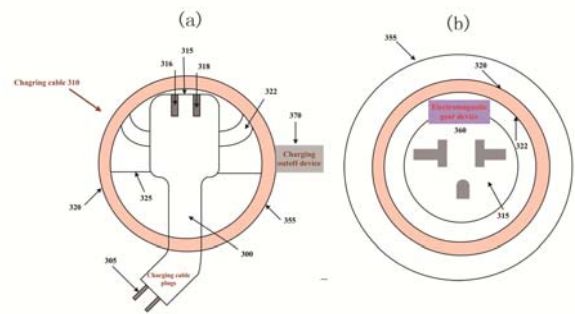

Fig. 3 (a) and (b) are the side profile and cross-section diagram of charging cable, respectively, wherein the cable port is connected with the electromagnetic gear (360).

Fig. 3(a) shows that the side section view of charging cable, with cable port 305 to be inserted into any sockets as described in Fig.2. In practice, charging cable 300 is DC power cable, whose port 305 can be linked to socket 205 in Fig.2. The charging cable also includes the insert port 315, composing the metal spring piece 316. Metal spring piece in total device is tightly connected with the electrical contract metal sheet in charging socket 205 to realize the electric conduction function. Socket port 315 contains the magnetic medium 318, which can be interacted with that corresponding in socket 200 by force of adsorption or rejection. In actual charging process, damage on vehicle charging device will take place when users pull the cable out of the charging socket. To escape from that, the charging cable 310 around the insulating layer is improved in its current structures. The surrounding insulating layer is composed of ball type rubber protective leather, the two sides of which is close to linking device 325 and socket 315. Furthermore, this protective leather can slide along the cross sectional surface 322, producing adsorption interaction between the magnetic medium at socket 315 and magnetic pole of charging cable 205. In reverse, when rejection interaction taking place between the magnetic medium at socket 315 and magnetic pole of charging cable 205, the cable will automatically fall off from the charging socket. This design is useful to prevent physical damage on charging device, as long as the users move vehicle but forget to pull the charging cable out.

Fig. 3(b) gives the cross-sectional profile of the charging system, including ball-type insulating section layer 320 and electric socket 315. 355 in Fig. 3(b) is the maximum radius of protection of the insulating layer. Based on the above implementation method, the charging system also includes the electromagnetic gear 360. In design process, there is an electrical contact metal sheet 201 at socket 205, and the charging plug port 315 can be connected with electrical contact metal sheet 201 for charging. In the non-charge state, the electromagnet at charging socket 315 and electrical contact 201 are the same polarity, producing mutual repulsion. Therefore, we firstly present the electromagnetic transmission device, which can be useful to real-time control the charging power behavior over connecting socket 315 with electrical contact 201. Electromagnetic gear 360 can be controlled

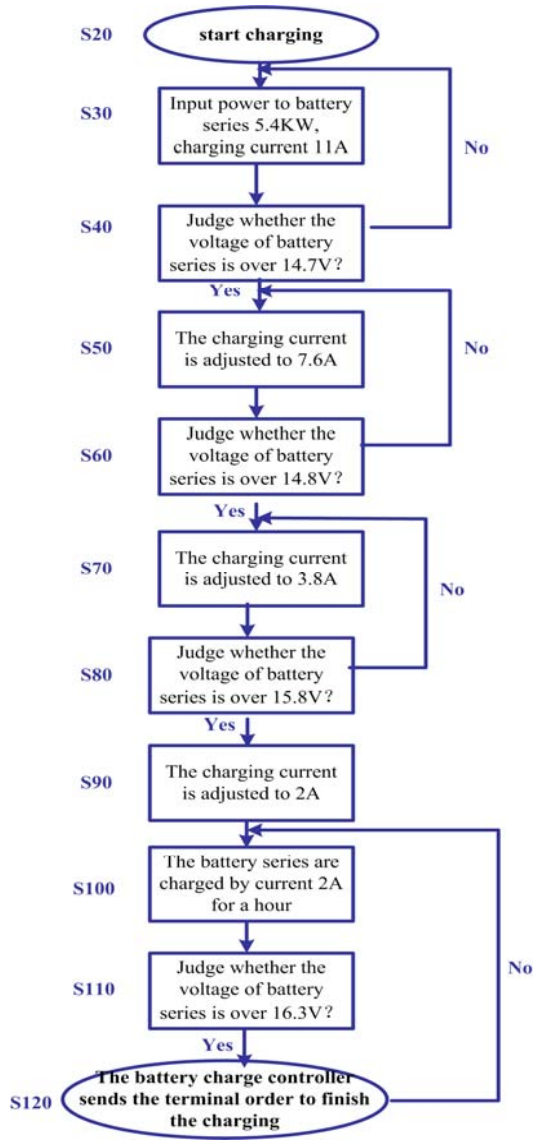

Fig. 2 The schematic flow chart of charging steps of the multi battery series in the electric vehicle.

through electromagnetic switch 366, which can select magnetic pole by laying electromagnet 220.

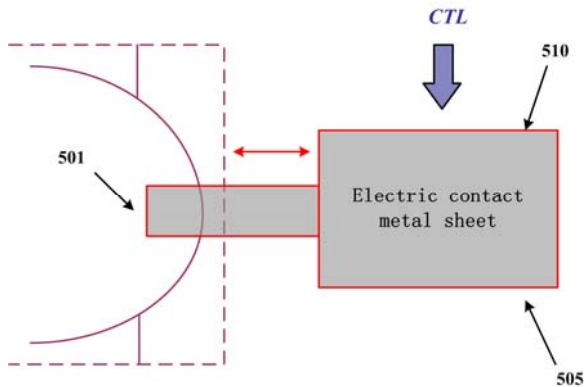

Fig. 4 (a) and (b) are the side profile and cross-section diagram of the socket rotating around the axis, respectively.

user interface through the automatic control device 150, which is composed of interlock circuit chip. Electromagnetic dive mechanism 360 executes the instruction and connects the cable tightly with electromagnet 220 covered with metal thin layer 318. When the vehicle is in the start state, controlling device 150 sends a start command to the interface, and at the same time it controls the interlocking circuit to realize the unlocking function which can cause opposite polarity between the electromagnet and metal thin layer 318, the magnetic rejection force will make the cable port and the charging socket detached, and finally finishing the whole charging process. The charging device also contains the charging off controller 370, which can be placed at outer of insulating protective cortex 320. As a modification design model to improve the quality of the efficiency, Li et al. have utilized Controller 370 for ending the charging state through controlling the electromagnetic gear to make the charging cable plug slip out of the socket [19]. By following the same procedure, we will also do the same thing to improve the charging efficiency of the socket. When users drive the car but forget to pull the cable out, charging off controller 370 can make the cable and socket automatically out, preventing physical damage on cable or socket.

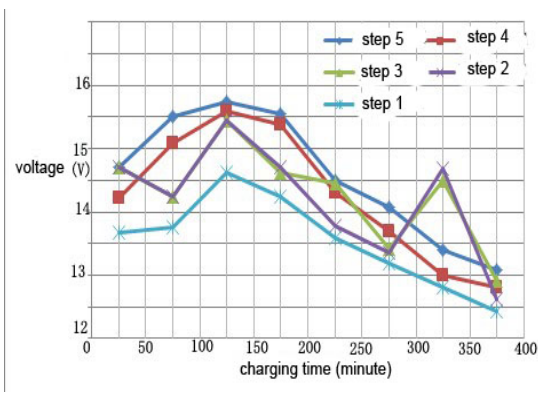

Fig. 5 The schematic diagram of the side section of the electrical contact metal sheet in charging socket.

According to the above implementation, further transformation on cable port 140 is realized in our design progress, that is, the port enters the charging socket in a certain inclination angle, which prevents the cable from falling from the socket caused by external force in the car charging process. When cable port 140 connects with charging socket 134, car backing up will cause the physical distortion of socket or even damage. To solve this problem, we propose the sliding socket 400 as it can seen in Fig. 4, which includes ball-type insulating plastic sheet rotating around the horizontal shaft and inclined shaft at the same time. This plastic sheet is used for connecting cross-sectional electrical contact 405. The size of the plastic sheet matches with the interface shape of socket 410. Furthermore, the socket in 405 can slide in the cross range. As it can seen in Fig. 4(a), the socket can slide freely along any angle or direction in the cross range. Socket 400 in Fig. 4 can be connected with curved cable 420 in Fig. 4(b), which can also be connected with charging sockets in different part of car as Fig. 1. Arc charging protective layer 410 contains arc aperture 411 and the electric contact chip 405 which can rotate around the shaft. The size and shape of 411 and 405 determine the space range of electrical contact chip rotating around shaft. Through the design of this kind of device, users can further control the angle connecting cable port and the charging socket by rotating the electrical contact chip. As a result, escaping from physical damage on cable and charging socket resulting from the moving in charging process. This result is comparable to that proposed by Li et al. Through such comparison work, we verify the correctness of Li's theorem and flexibly apply to our studying case [19].

Fig. 5 is the schematic diagram of automatic charge cut-off device with electromagnet. 501 is the conductor sheet metal, which is used for conductive charging cable input current. The whole charging process can be controlled by electromagnetic gear 505, which insets and pulls out the automatic control conductor mental piece by means of the single chip circuit. When the user stops electric vehicle and sends order to the electromagnetic gear for charging, the electromagnetic gear receives the order and sends control signal 510, and controls the conductor metal piece closing and entering the charging socket at the same time, at last, realizes the function of electric vehicle charging.

## IV. CONCLUSION

Comparing with the current advantages of electric vehicle charging power devices, merits of the method proposed in this paper can be listed as follows: firstly, magnetic medium is plated respectively in the cable port and charging socket contact to realize the charging and power function according to the needs of users; secondly, the electrical contact metal sheet at charging socket can shift under the action of electromagnetic gear to realize the select of vehicle charging and power down mode; finally, Charging cable port embedded with arc insulation plastic protective layer can prevent leakage and short circuit of charging circuit. This protective layer is located at the gap of cable and socket contact, adjusting the gap distance and the electrical contact pole distribution to achieve automatic adsorption and releasing of socket and charging cable.